\documentclass[aps,prl,twocolumn]{revtex4}

\usepackage{epsfig}
\usepackage{graphicx}  
\usepackage{bm}        
\usepackage{amssymb}   

\begin{document}

\title{ Carrier drift velocity and edge magnetoplasmons in graphene }

\author{ I. Petkovi\'{c}$^{1,2}$$^*$, F.I.B. Williams$^{1,3}$,
K. Bennaceur$^{1,4}$, F. Portier$^{1}$, P. Roche$^{1}$ and D.C. Glattli$^{1,5}$}

\affiliation{
 $^{1}$  Service de Physique de l'\'{E}tat Condens\'{e}, Commissariat \`{a} l'\'{E}nergie Atomique, 91191 Gif-sur-Yvette, France.\\
 $^{2}$ Laboratoire National de M\'{e}trologie et d'Essais, 29 avenue Roger Hennequin, 78197 Trappes, France.\\
 $^{3}$ Institute for Solid State Physics and Optics, Wigner Research Centre for Physics, P.O. Box 49, H-1525 Budapest, Hungary\\
    }

\altaffiliation {
     $^{4}$Now at Department of Physics, McGill University, Montreal, H3A 2T8, Canada.\\
 $^{5}$Also at \'{E}cole Normale Sup\'{e}rieure, Laboratoire Pierre Aigrain, 24 rue Lhomond, 75231 Paris Cedex 05, France.}

\begin{abstract}
We investigate electron dynamics at the graphene edge by studying the propagation of collective edge magnetoplasmon (EMP) excitations. By timing the travel of narrow wave-packets on picosecond time scales around exfoliated samples, we find chiral propagation with low attenuation at a velocity which is quantized on Hall plateaus. We extract the carrier drift contribution from the EMP propagation and find it to be slightly less than the Fermi velocity, as expected for an abrupt edge. We also extract the characteristic length for Coulomb interaction at the edge and find it to be smaller than for soft, depletion edge systems.
\end{abstract}

\pacs{}
\maketitle

\email{ ivana.petkovic@cea.fr}
The Quantum Hall Effect (QHE) occurring in 2-Dimensional Electron Systems (2DES) relies vitally on edges, both to accumulate  charge and to support non-dissipative chiral currents. The time dependent Hall effect takes the form of Edge Magneto-Plasmons (EMP), quasi one dimensional gapless elementary excitations which are split off from the bulk magneto-plasmon modes by the sample boundary\cite{Fetter85,Mast85,Glattli85,Volkov85,Volkov86,Volkov88,Aleiner92,Allen83,Andrei88,Ashoori92,Zhitenev93,Ernst96,Kumada11}.  Ever more closely confined to the edge as frequency and wavevector are increased, they are a tool of choice to investigate edge structure and dynamics. Seen in both classical\cite{Mast85,Glattli85} and quantum inertial 2DES\cite{Allen83,Andrei88,Ashoori92,Zhitenev93,Ernst96,Kumada11}, they could also exist for electrons in graphene despite the very different gapless Dirac dynamics\cite{balev,wang}. We present experiments which establish firstly that EMP do indeed exist in graphene and then, from the propagation properties, we extract both the velocity of the chiral current on the edge and the characteristic length for edge Coulomb interaction.

Electrons in graphene obey two-dimensional massless relativistic-type dynamics with speed $v_{F}\sim 10^8$  cm s$^{-1}$ (Fermi velocity). The honeycomb lattice structure adds pseudospin-orbit coupling. Energy dispersion  is linear and gapless around the band crossing ``Dirac" points situated at the Fermi level for charge neutrality\cite{Wallace47}. Charging sweeps the Fermi level through the crossing to make a smooth transition from electron to hole behaviour. The electronic edge is then never defined by depletion (the ``soft" edge of the usual gapped 2DES) and the electrons should always feel the abrupt work function potential of the sample edge. On approaching the boundary in perpendicular magnetic field the energies of the edge Landau states increase because of confinement and eventually  emerge above the Fermi level. The resulting confinement force is balanced by carrier drift along the edge which constitutes a  chiral current with vanishing backscattering, the essential ingredient for a quantum Hall effect which, in graphene, because of the unusual degeneracies induced by the pseudo-spin coupling, displays an anomalous integer filling factor sequence $\nu=2,6,10,...$
\cite{Novoselov05,Zhang05}, $\nu$ being the number of electrons per flux quantum.

Classically EMP appear as propagating periodic variations of charge excess/deficit on the edge which, by creating an electric field, induce drift currents which redeposit the charge to advance the wave along the boundary. Because the charge accumulation is proportional to the Hall drift current, the phase velocity is determined by the Hall conductivity $\sigma_{xy}$ and electrostatics. On the edge the charge drifts at velocity $v_D$, pulling the electrostatic field with it and thereby adding $v_D$ to the Hall conductivity propagation. In quantum language EMP become a propagating local variation of chemical potential (filling of the Landau level as it emerges above the Fermi level at the edge) which drives a local Hall current which by continuity is accompanied by charge propagation\cite{macdo}. The inward force driving the electron drift is set by the variation of chemical potential with charge imbalance and consists of two parts: the slope of the emerging Landau level and the electrostatic field from the imbalance which give rise respectively to the $v_D$  and $\sigma _{xy}$ terms in the dispersion relation \cite{macdo,mikhi}
\begin{equation}
\omega = v_{\varphi}\,q=\left[\frac{2\,\sigma_{xy}}{\varepsilon_{\rm eff}}\left( \ln\frac{2}{\left\vert
\,q\right\vert\! w}+C \right)+ v_D\right]q,
\label{eq1}
\end{equation}
\noindent where $\varepsilon_{\rm eff}$ is the effective dielectric substrate screening constant, $q$ the wave vector, $w$ a lower cut off length in the electrostatics which can be expected to be of the order of the magnetic length $\ell_{B}=\left(\hbar c/eB\right)^{1/2}$, but could also result from lateral spread of the charge imbalance. The constant $C=1$ for a step function density profile, but in general it is a function of the equilibrium profile at the edge\cite{Volkov88}.
The basic form of Eq. (1) is the same for graphene as for the usual 2DES because Lorentz force drift motion depends ultimately only on the Lorentz transformation to the zero electric force frame for the electrons\cite{jackson}. The slope of the Landau level energy, and therefore the drift velocity, is, however, dynamics dependent and scales with the respective Fermi velocities. The density profile is determined principally by electrostatics with narrow incompressible strips where the density situates the Fermi level in the Landau level gaps\cite{scglazman}. Electrostatics for a charged equipotential sheet imposes the limiting form of the edge density profile to be $n_s (x)\sim x^{-1/2}$ where $x$ denotes distance from the edge, as opposed to $n_s (x)\sim x^{1/2}$ for a depletion edge. The repulsive force responsible for the drift velocity results from the increase in energy $W_{n}\sim p\, v_F\sim \sqrt{n}\hbar v_{F}/\ell_B$ of the Landau function of level index $n=(\nu - 2)/4$ as it is squeezed against the hard edge: in graphene, the effective confining electric field is $E^{\rm eff}_{n}=-(1/e)\delta W_{n}/\delta x \sim -\hbar v_{F}/e\ell_B ^2$, where $\ell_B$ characterises the width of the wavefunction. Electrons drift along the edge in this potential gradient at velocity $v_{D} = cE_{\rm eff}/B \sim v_{F}$. For graphene the drift velocity term in Eq. (1) can account for up to $\sim 50\%$ of the total and a measurement of EMP velocity offers an excellent probe of this important but hard to access quantity.

The recent interest in bulk magnetoplasmons in graphene\cite{roldan,Crassee,Yan} has led to experiments on magnetic field splitting of the bulk plasmon frequency in the far-infrared absorption spectrum of spatially modulated graphene\cite{Crassee,Yan}. The absorption resonance which was found to decrease in energy with increasing field is expected to evolve into an edge-localised mode. However the propagation velocity, which is the central feature of EMPs, was not accessible in those experiments.

In the present experiment, we create a wavepacket of EMPs with a voltage pulse and measure the time of propagation between two points on the edge such that left and right perimeter paths are different. Combining propagation time with path length measured by optical microscopy affords a very direct measurement both of sign and amplitude of the group velocity which, from equation (1), is expected to be

\begin{equation}
v_{g}=\frac{d\omega}{dq}=\frac{2\,\sigma_{xy}}{\varepsilon_{\rm eff}}\left( \ln\frac{2}{\left\vert
\,q\right\vert w}+C-1\right)+v_{D}.
\label{eq2}
\end{equation}

\begin{figure}[t]
\vspace{2mm} \centerline{\hbox{
\epsfig{figure=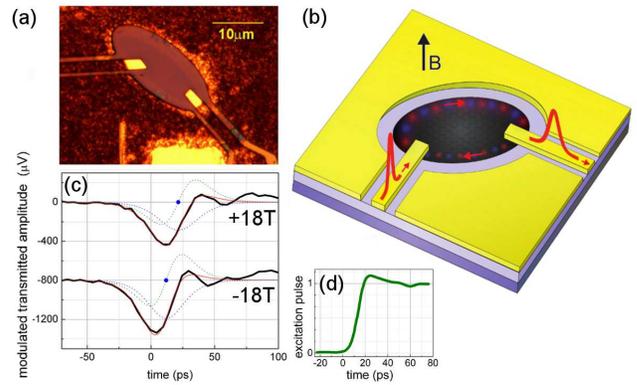,width=85mm}}}
\caption{(a) Optical photograph of graphene sample with edge coupling electrodes terminating the coplanar waveguides. (b) Measurement configuration: EMP wavepacket excited by 7 or 11 ps rise time 100 mV step function propagates along edge (white arrows). (c) Demodulated responses, offset for clarity, at $B=\pm18$ T (filling factor $\nu =2$) constructed by subtraction of waveforms at $\pm$15V side gate potential for 11ps excitation. The arrival time difference arises from unequal left and right path lengths for oppositely directed magnetic fields. Overlay in red is best fit to weighted sum of first (blue dots) and second (solid blue) derivatives of a Fermi function (see text). Blue dot is at fitted arrival time: 21.4 ps for +18T and 11ps for -18T. (d) Direct measurement of 7 ps rise time excitation pulse broadened to 13 ps by the receiver amplifier.}
\label{uninset}
\end{figure}

Broadband microwave transmission was measured between 2 points on the periphery of an oval shaped graphene flake of about $15\times 5\mu$m exfoliated from natural flake graphite onto the surface of a 290nm thick SiO$_{2}$ layer commercially grown on an insulating Si wafer as shown in Fig. 1(a,b). The flake is separated by $\sim 3 \mu$m  from the surrounding ground plane common to  the two 50 GHz bandwidth microwave coplanar waveguides whose $2 \mu$m wide centre conductors overlap the edge by about $2 \mu$m to serve as probe electrodes. One is capacitive and the other ohmic and they divide the perimeter in a ratio of about 2:1. The waveguides are tapered out to connect to a coplanar microwave printed circuit board placed horizontally and connectorised to vertical coaxial lines leading to the room temperature excitation and signal analysis instruments. Total attenuation at 50GHz was -25dB. The sample was identified as graphene, and its quality monitored throughout fabrication, by scanning Raman spectroscopy (Supplementary Fig. S1). It was placed in cryogenic vacuum at 2.2K in a vertical magnetic field of up to $\pm 19$T where it was annealed and its density adjusted by heating to 425 K. Carrier density was identified from Shubnikov-de Haas oscillations in microwave transmission at $\sim{5}$ GHz, a technique analogous to a 2-terminal DC measurement, but mobility was estimated from DC measurements on other, similarly prepared samples\cite{Bennaceur}. For the time of flight experiment pulse excitation and analysis are performed by a dual-channel time domain reflectometry module in transmission mode. A fast 11 or 7 ps rise time step function pulse of $\sim100$ mV (see Fig. 1(d)) is sent to one electrode to excite a wavepacket which propagates to the other electrode where it induces the signal sent to the receiver through a low noise 0.1-65 GHz preamplifier and digitally recorded by the picosecond sampling oscilloscope. The EMP wavepacket voltage is differentiated by the capacitive coupling to give an approximately symmetric output pulse, but the raw waveform is a superposition of the response of the sample with the response of the transmission lines and amplifiers. To select the sample contribution, and to reduce noise from amplifier drift, a one hertz sidegate modulation voltage $V_{sg}=\pm15$V is applied to the ohmic contact electrode \textit{via} a broadband bias tee. The 
two signal waveforms are subtracted from one another to eliminate the response of all but the sample, it being the only element sensitive to $V_{sg}$. The response was identical upon interchange of input and output lines with simultaneous reversal of magnetic field.

\begin{figure}[t]
\vspace{2mm} \centerline{\hbox{
\epsfig{figure=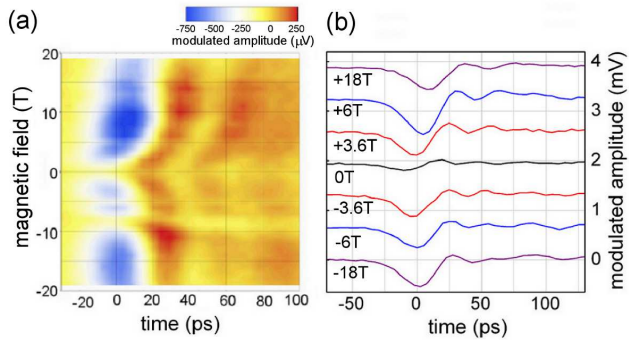,width=85mm}}}
\caption{(a) 3D colour plot of modulated signal amplitude in the propagation time - magnetic field plane. (b) Five signals chosen at filling factors $\nu=2,6,10,$ offset for clarity. From top to bottom: 18T, 6T, 3.6T, 0T, -3.6T, -6T, -18T. These data correspond to $11$ps excitation and 5ps sampling resolution.}
\label{uninset}
\end{figure}

The physics resides in the wave packet propagation times. Fig. 1(c) shows arrival signals for $\pm$18T\ fields \emph{vs} time measured from the leading edge of the excitation at the source module while Fig. 2 paints the overall picture of arrival time profiles for magnetic fields $-19<B<+19$ T corresponding to Landau level filling factors $\nu=n_s h c/eB \gtrsim 1.9$, where $n_s \simeq 0.87\times {10^{12}}$cm$^{-2}$ is the surface carrier density. Detailed analysis of the data proceeds by remarking that the side gate potential alters the filling factor at the edge, influencing both velocity and attenuation. Variation of attenuation modulates amplitude and replicates the basic form of the wavepacket whereas variation of velocity modulates arrival time and differentiates the basic shape. The mean arrival times should be the same, motivating a fit  to a weighted sum of symmetric and anti-symmetric parts derived from a single function. Guided by the exponential shoulders of the excitation step, we chose a Fermi function for which a typical fit is illustrated in Fig. 1(c). Fig. 3 plots arrival times deduced from the fits as a function of oppositely oriented magnetic fields. Field differentiation of the data calculated by subtracting arrival pulse waveforms at neighbouring fields corroborates the side gate modulated results (see Supplemental Material).

The difference in arrival times for $\pm\left|B\right|$ is compatible with the left and right edge path lengths of 14 and 26 $\mu$m around the sample from emitter to receiver for fields above about 2.5 T. Converting arrival time into propagation time, however, requires knowing the time of emission. We approach this problem in three ways. Firstly, the ratio of propagation lengths allows us  to deduce time zero from the two arrival times supposing that the left and right speeds are the same at the same field intensity and that the wave propagates along the perimeter. Secondly, as $B\rightarrow0$ the propagation velocity $v_g \propto\sigma_{xy}\propto1/B\rightarrow\infty$ will be limited by the change in nature of the wave towards a zero field, non-chiral bulk plasmon, propagating with group velocity
$v_{g}(B=0)=\sqrt{\frac{k_F}{q}v_F \frac{e^2}{h}\frac{\pi}{\varepsilon_{\rm eff}}}\approx \sqrt{\frac{k_F}{q}v_F v_{\rm g}(\nu =2)}$,
where $q$ is of the order of the inverse width of the wave packet and $k_F=\sqrt{\pi n_s}$ is the Fermi wavevector\cite{das Sarma}. The arrival time is then the time required to travel at this much faster velocity along the direct internal path between emitter and receiver electrodes. Thirdly, if we neglect the formation of the bulk plasmon, the arrival times should extrapolate linearly to zero for $B \rightarrow 0$. The three methods agree to $\pm 1$ps and fix the zero of the time scale on Fig. 3 from which we calculate the propagation velocities.

\begin{figure}[t]
\vspace{2mm} \centerline{\hbox{
\epsfig{figure=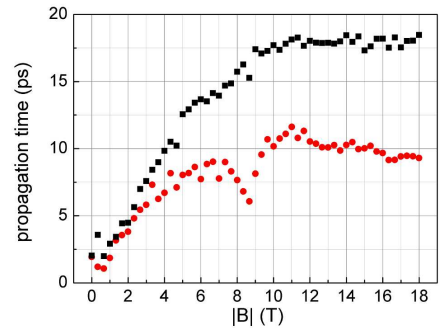,width=60mm}}}
\caption{Propagation times between emitter and receiver structures as function of perpendicular magnetic field for opposing orientations. The determination of zero time is described in the text. Filling factors $\nu=10,6$ and 2 are at 3.6, 6 and 18T respectively. Black (red) symbols correspond to the positive (negative) sign of the magnetic field. These data correspond to $7$ps excitation and 2ps sampling resolution.}
\label{uninset}
\end{figure}

Three significant features emerge from the propagation time data of Fig. 3. One is the positive, quantitative identification of  chirality at high field, since the propagation times are proportional to left and right path lengths for oppositely directed fields. Another is the unique pulse response at fixed field orientation indicating that only a single path is taken and no counter-propagating mode is detected. The third is that the propagation times and therefore velocities exhibit a plateau structure in magnetic field with the same filling factor sequence as the Hall conductivity. Filling factors $\nu=10,6$ and 2 are estimated to be at 3.6, 6 and 18T respectively. For the longer path the quantization is well developed at $\nu=2$ and is clearly discernable at $\nu=6$. On the shorter path the quantization for $\nu=2$ is observed, but for $\nu=6$ it is somewhat blurred by the appearance of the dip feature between the two plateaus, the origin of which is not understood.

The propagation velocities are determined by combining the propagation time results of Fig. 3 with the perimeter path lengths (as shown as function of the filling factor in Figure 3 in the Supplemental Material). We extract the drift velocity $v_D(\nu)$ from the experimental EMP propagation velocity at filling factors $\nu=2,6$ and 10 using only the functional form in filling factor dependence of the two contributions - closely linear dependence for the Hall conductivity and the theoretically calculated ratios at mid gap chemical potential for the drift velocity\cite{Abanin0607,Brey06,Peres06,Delplace10}. The drift velocity so extracted is then independent of the fitting parameter $qw$. We find $v_D=(0.7\pm 0.3) \times 10^8$ cm s$^{-1}$ for $\nu=2$, very close to the theoretical estimate $v_D\simeq0.7 v_F$ for $\nu=2$ at a sharp edge.

We estimate the product $qw$ from the part of the propagation velocity linear in $\sigma_{xy}$ by virtue of the logarithmic factor in equation (\ref{eq2}). To extract $w$ from $qw$ we estimate $q$ by converting temporal width $\Delta \tau$ given by the excitation rise time into spatial width $\Delta s=v_g \Delta \tau\simeq {1/q}$. The received pulse widths are $\Delta\tau_{p} \sim 25$ps of which about $ 20$ ps are due to the finite rise time of the excitation pulse combined with the receiver bandwidth. Replacing the 11ps with a 7ps excitation pulse made only a marginal difference of about 2ps to the received pulse width. Setting $C=1$ in equation (2), strictly only appropriate for a step function edge profile, leads to $w=500 \pm 200$ nm, much greater than the expected $\ell_{B} \sim 6$ nm. It could possibly be interpreted as lateral width of charge accumulation, but it could also result from not taking into account the $x^{-1/2}$ density divergence on approach to the edge which, classically, accumulates charge ($\dot{n} \simeq -\textbf{v.}\nabla n$) of opposite sign to that of the edge itself lending a dipolar nature to the mode.

EMP attenuation can be estimated from the ratio of response amplitudes for the two unequal paths if the attenuation coefficient is supposed the same for both. From the arrival signals for $\nu =2$ shown in Fig. 1 the attenuation length is $70\pm{30}\,\mu$m corresponding to a relaxation time of $50\pm{20}$ ps, three orders of magnitude longer than the $\tau \sim 0.05$ ps Drude relaxation time applicable for bulk plasmons in similar samples\cite{Bennaceur} which showed mobilities of $\sim 5000$cm$^2$s$^{-1}$V$^{-1}$.

The propagation times for the two paths are no longer experimentally distinguishable below about 2.5$\,$T ($\nu\gtrsim 14$), indicating that EMP excitations are no longer clearly chiral. From our experience on transport measurements in similar samples, these are fields for which full chirality is no longer expected to be preserved as disorder supresses the QHE gaps\cite{Bennaceur}. Any echo from propagation around a full circle is at least 20 $\times$ smaller than the single path signal,  indicating that at least one contact pad almost fully absorbs the wavepacket. For an ohmic contact, the low (50$\Omega$) impedance detector would effectively short-circuit the EMP wave impedance ($\rho_{xy}\sim 10^4 \,\Omega$) and suppress re-emission. On the other hand, the second contact pad is capacitive because the main pulse has the shape of the derivative of the injected step voltage and there is no ohmic connection between the two edge electrodes. A model for chiral wave absorption by contact shunts is presented in Supplemental Material.

In summary, graphene EMP differ from their inertial gapped 2DES counterparts not by the basic form of the dispersion relation but by the effect of the hard edge and gapless relativistic dynamics on the drift velocity. We identify them in graphene by observing chirality, quantization of velocity and low attenuation. We see no evidence of counter-propagating states under our experimental conditions, but these are far from the neutrality point for which such modes were proposed\cite{Abanin}.
The EMP propagation velocity exceeds the Fermi velocity, affirming the collective nature of the excitation. The carrier drift component $v_{D}$ is less, but comparable, to the Fermi velocity, in keeping with gapless, massless particle behaviour at a sharp edge. The fit to a characteristic length of 500 nm in the edge Coulomb interaction is one order of magnitude smaller than  observed in soft edged GaAs\cite{Kumada11} but it is nearly two orders of magnitude greater than the magnetic length. Although this could reflect lateral charge distribution, it could also result from the oversimplified model which does not take into account the divergence of the edge density profile or electrostatic screening by the compressible strip.

For use in plasmonics, bulk plasmons in graphene offer great advantages over metals in being electrically tunable and in having long relaxation times \cite{Koppens11,Fei11,Ju11,Koppens12,Basov12}. Edge magneto-plasmons offer further features over bulk plasmons: they are $>1000 \times$ less damped, they are chiral with chirality determined by gate voltage sign and the propagation velocity is linearly rather fourth root dependent on gate voltage. Graphene EMP could extend considerably the frequency domain of graphene based plasmonics and open up a whole new domain of chiral plasmonics.

In the course of completing the work reported here, we became aware of similar work on EMP propagation in mm size samples of graphene on SiC involving ns rather than ps propagation times\cite{Kumada12}.

We gratefully acknowledge discussions with M.B\"{u}ttiker, L.I.Glazman, M.Goerbig, J.-N. Fuchs, E.Andrei, G.Li and I.Skachko. We thank P.Jacques, P.-F.Orfila, C.Chaleil, M.de Combarieu, P.Forget and P.Pari for technical advice and assistance. This work was supported by the ERC advanced Grant MeQuaNo No. 228273 and the RTRA Gamet Grant No. 2010-083T.

\onecolumngrid
\begin{center}
\Large{Supplemental material: Carrier drift velocity and edge magnetoplasmons in graphene}
\end{center}


\normalsize

\vspace{4mm}
\section*{Sample Fabrication}

\begin{figure}[h!]
\vspace{2mm} \centerline{\hbox{
\epsfig{figure=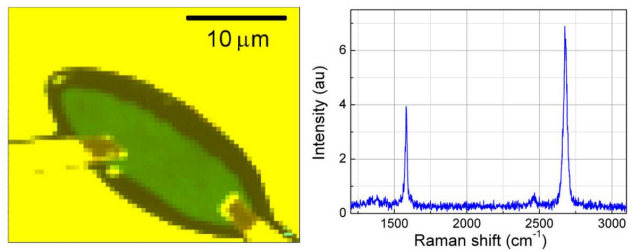,width=85mm}}}
\caption{\textbf{Raman sample characterisation.} (a) Raman map of the sample, where green corresponds to graphene. (b) Raman spectrum in the middle of the graphene sample.}
\label{raman}
\end{figure}

 The sample was made by exfoliating graphene from natural Indian graphite procured from NGS Naturgraphit GmbH. It was deposited by the standard Scotch tape procedure on a 290nm thick layer of thermal SiO$_2$ commercially grown on an insulating Si wafer of room temperature resistivity $\sim 8$k$\Omega$-cm. The wafer was prepared in a Piranha solution ($\rm H_2SO_4+H_2O_2$) followed by an oxygen plasma etch. Spurious grains of graphite were removed with high pressure oxygen plasma etching, during which the graphene was protected with a PMMA/HSQ mask. The coplanar waveguide (CPWG) conductors of Ti(5nm)/Au(200nm) were then deposited by evaporation in high vacuum through an electron beam patterned PMMA mask. The central conductors extend onto the edge of the graphene by about 2$\mu$m as seen in the photo of Fig.1(a) of the main text. After liftoff the graphene sample was characterised with scanning micro-Raman spectroscopy (see Figure \ref{raman}) with a 532nm wavelength laser and found to be a monolayer with low impurity content. The wafer was cut into a $3\times 3$mm$^2$ flip chip so as to leave the sample centrally placed and for the on chip CPWG to make contact with the CPWG of the microwave printed circuit board (PCB).

\vspace{-2mm}
\section*{Experimental configuration}

The sample was mounted perpendicular to the (vertical) field axis in a hermetic insert made for the bore of the 19T superconducting magnet. The magnet was cooled in a $^4$He cryostat with lambda plate cooling of the coil region to $\sim 2.2$K. The sample was annealed by heating its holder to $150^{\circ}$C for two hours while maintaining a cryogenic vacuum with the cold 4.2K walls of the insert. Transmission measurements of $S_{21}$ as a function of field at a few GHz allowed us to estimate the electronic density to be about $0.87\times 10^{12}$cm$^{-2}$.

\vspace{-2mm}
\section*{Design of the microwave circuit}

 We designed the microwave circuit to and from the sample to have 50GHz bandwidth with the help of CST Microwave Studio software.
Two coaxial cryogenic cables connect two V-connector hermetic feed-throughs from the top of the insert cryostat to the horizontal sample holder PCB assembly  \emph{via} Mini-SMP right angle connectors modified to have less dielectric to ensure reduced reflection and with a bent central conductor to make  solderless connections to the tapered coplanar waveguides (CPWG). The waveguides are etched onto a 17 $\mu$m thick gold-on-copper face of a TMM10 substrate PCB with a dense array of via holes. The waveguides terminate at contact pads on adjacent edges of a $2\times 2$mm$^2$ square hole over which is placed the $3\times 3$mm$^2$ chip from an undoped Si wafer containing the graphene sample. We used undoped Si to limit dielectric losses. The ``flip chip'', so called because it is mounted face down on the PCB to ensure contact between its own CPWG and those of the PCB, contains sample specific tapered waveguides which terminate at the graphene in either an ohmic or capacitive contact. The flip-chip is held in place by an optically flat metallic backing plate attached to a bent beam spring of Vespel plastic. All contacts to the PCB are solderless to resist the high 450-500K temperatures required for annealing, care being taken to ensure sufficient elasticity of the contacts to compensate thermal contraction. The frequency-dependent loss in the PCB and flip-chip CPWG was measured by the transmission coefficient $S_{21}$ between the two Mini-SMP connectors with a continuous CPWG on the flipchip. It has a linear baseline with small oscillations in frequency, and is about -7dB at 50GHz. The insertion loss of the total circuit including 4m of cryogenic coaxial cable is around 25dB at 50GHz. The complete circuit is operational up to 50GHz. At higher frequency vertical standing wave modes start to appear across the thickness of the PCB, leading to substantial loss. The details of the microwave setup will be published elsewhere.

\vspace{-2mm}
\section*{Magnetic field derivative of the transmitted signal}

 \begin{figure}[h!]
\vspace{2mm} \centerline{\hbox{
\epsfig{figure=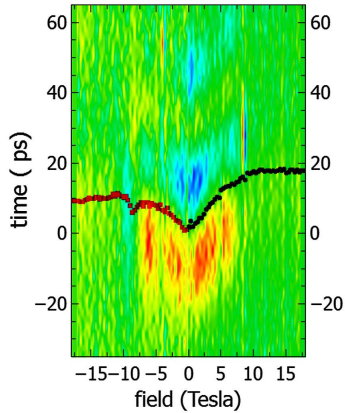,height=55mm}}}
\caption{\textbf{Field derivative of the transmitted signal.} Rainbow colour plot of the derivative of the transmitted signal with respect to the magnetic field. Positive to negative values range across the spectrum from red to blue. The black and red superimposed data points are the transmitted pulse arrival times as shown in Figure 3 of the main text, where the signal is modulated with respect to gate voltage.}
\label{dSdB}
\end{figure}

The derivative with respect to field of the arrival signal is illustrated by the  rainbow colour plot in Figure \ref{dSdB} which is obtained by subtracting consecutive 0.33T interval direct (unmodulated) transmission signals. The propagation times deduced from the analysis of the derivative with respect to gate voltage (Figure 3 in the main text) have been superimposed to facilitate comparison from which we conclude that the derivative in field gives similar pulse propagation times to those obtained from the derivative with respect to gate voltage. However, unlike the gate voltage modulated signal, the field derivative is more subject to experimental drift as the difference is made between acquisitions separated by long times and shows inferior signal to noise.

\vspace{-2mm}
\section*{Edge magnetoplasmon propagation velocity}

\begin{figure}[hh!!!]
\vspace{2mm} \centerline{\hbox{
\epsfig{figure=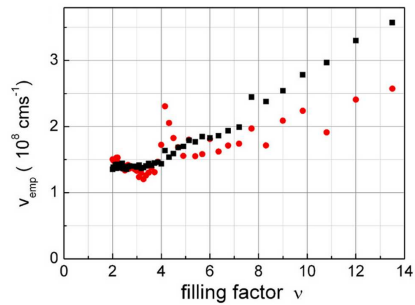,width=55mm}}}
\caption{\textbf{Propagation velocity as function of the filling factor.} Propagation velocity as function of the filling factor for both signs of the magnetic field from the data presented on main text Figure 3 with the same colour coding.}
\label{FigS3}
\end{figure}

\vspace{-2mm}
\section*{Modeling the ac response of the graphene sample}

Below we derive the ac response of a graphene sample in the idealized case of perfectly transmitting edge channels and
assuming that, at a given filling factor, all the chiral edge channels running in parallel propagate with the
same velocity. The aim is to understand the effect of a perfect ohmic contact connected to an external circuit of negligible impedance on the chiral EMP propagation. Does it fully absorb the EMP chiral wave? How is the pulse response modified when the magnetic field is reversed?

To model the ac response, we start from the gauge invariant, current conserving approach developed in \cite{SM1} which is particularly appropriate for chiral quantum Hall edge channels. We first derive the ac conductance for two simple situations: the case of two ohmic contacts and the case of one ohmic and one capacitive contact. The results of \cite{SM1} are valid in the low frequency limit where the long range interaction can be simply taken into account by introducing capacitances to ground. In the second part, we extend our analysis to include the
high frequency regime relevant to edge magneto-plasmons by adapting the approach of \cite{SM2}
for a 1D wire with short range interactions to the chiral case. The assumption of chirality leads to a dispersion relation for the magneto-plasmon mode where the electron drift velocity simply
adds to the plasmon velocity with short range interactions. Generalization to long range interactions would lead directly to Eq.(1) of the main text. Regarding the ac current response to a voltage pulse, our calculation shows that the only relevant modification due to interactions is to change the propagation velocity, while the absorptive condition of the contact remains unchanged. In particular the existence of at least one ohmic contact is shown to prevent the formation of pulse echoes and would seem responsible for our observing only a single pulse in the current response.

\subsubsection{I. low frequency regime}
\vspace{4mm}

\begin{figure}[h]
\vspace{2mm} \centerline{\hbox{
\epsfig{figure=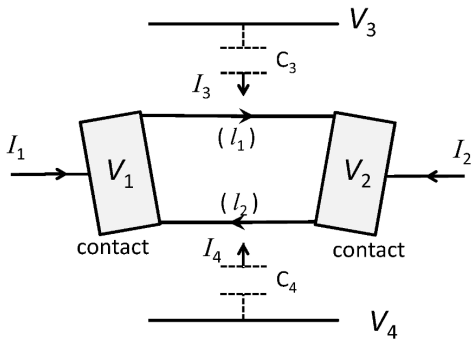,height=50mm}}}
\caption{\textbf{Two ohmic contacts.} 
Conduction paths between two ohmic contacts connected by two chiral states of lengths $l_1$ and $l_2$. }
\label{FigS1}
\end{figure}

 We first consider the case of two ohmic contacts as shown in Figure \ref{FigS1}. The lengths of the filling factor $\nu$ chiral edge states connecting contacts (1) and (2) are $l_{1}$ and $l_{2}$. In the limit where one can linearize the electron dispersion relation around the chemical potential of the edges, the ac transport Kirchhoff's laws relating the ac current $I_{\alpha}$ at the input of contact $\alpha$ to the ac voltage $V_{\beta}$ applied on contact $\beta$ are
\begin{equation}
I_{1}=\nu \frac{e^2}{h}\left(V_{1} - V_{2}e^{i\phi _{2}} - \widetilde{V_{4}}(1 - e^{i\phi _{2}})\right)
\end{equation}
\begin{equation}
I_{2}=\nu \frac{e^2}{h}\left(- V_{1}e^{i\phi _{1}} + V_{2}  - \widetilde{V_{3}}(1 - e^{i\phi _{1}})\right)
\end{equation}
\begin{equation}
I_{3}=\nu \frac{e^2}{h}\left(- V_{1}(1 - e^{i\phi _{1}})  +\widetilde{V_{3}}(1 - e^{i\phi _{1}})\right)
\end{equation}
\begin{equation}
I_{4}=\nu \frac{e^2}{h}\left(- V_{2}(1 - e^{i\phi _{2}})  +\widetilde{V_{4}}(1- e^{i\phi _{2}})\right)
\end{equation}
where $\phi_{1(2)}=\omega l_{1(2)}/v_{D}=i\omega\tau_{1(2)}$ is the temporal phase, $\omega$ the frequency, and $v_{D}$ the carrier drift velocity. In the above set of equations the terms $e^{i\phi _{i}}$ have to be replaced by $1+i\omega \tau_{i}-1/2(\omega \tau_{i})^2$ in the limit of small $\omega$. In this limit, the voltage $\widetilde{V}_{3(4)}$ represents a fictitious voltage drop $I_{3(4)}/(-i\omega C_{q_{3(4)}})$ across the quantum capacitance $C_{q_{3(4)}}=e^{2}\tau_{1(2)}/h$. Upon including the geometrical capacitance $C_{3(4)}$, it is related to the actual voltage source by $V_{3(4)}=\widetilde{V}_{3(4)}+ I_{3(4)}/(-i\omega C_{3(4)})$. 
In keeping with \cite{SM1} equations (1-4) are invariant under a global shift of all potentials by a common quantity $U$ and they show current conservation as they include the displacement currents $I_{3}$ and $I_{4}$.

In the absence of backscattering between upper and lower edges, the dynamics of each edge are decoupled. If the two contacts are not equally spaced along the perimeter, as is intentionally the case for our graphene sample, i.e. $l_{1}\neq l_{2}$, the ac response of the current is expected to change under magnetic field reversal. As an example, we take $V_{1}=V$ while $V_{\alpha \neq 1}=0$ and look at the current response $I=-I_{2}$. This would correspond experimentally to the situation where the input transmission line is connected to contact (1) and the output transmission line to contact (2) and the finite characteristic impedance of the transmission line $Z_{C}=50\Omega$ is neglected with respect to $h/\nu e^{2}$ (several k$\Omega$). To first order in  $\omega$
 \begin{equation}
I=\nu \frac{e^2}{h}V + i\omega \frac{C_{q_{3}}C_{3}}{C_{q_{3}}+C_{3}}V
\end{equation}
Reversing the sign of the magnetic field, and therefore the direction of propagation, leads to a similar expression but with index (3) replaced by the index (4). Thus, unless the upper and lower lengths are equal,  $I(B)\neq I(-B)$
\cite{Christen97}. Because of the displacement currents, the $B$ parity imposed by the Onsager-Casimir relations for two-terminal samples no longer holds at finite frequency.

In the limit of infinite geometrical capacitances $V_{3(4)}=\widetilde{V}_{3(4)}$, the Coulomb interaction is fully screened and the following expression for $I(V)$ is valid at any frequency
 \begin{equation}
I=\nu \frac{e^2}{h}V e^{i\phi_{1}}
\end{equation}
or in the time domain $I(t)=\nu \frac{e^2}{h}V(t-\tau_{1})$. If the magnetic field is reversed $\tau_{1}$ and $\phi_{1}$ are replaced by $\tau_{2}$ and $\phi_{2}$.

\vspace{10mm}

\begin{figure}[h]
\vspace{2mm} \centerline{\hbox{
\epsfig{figure=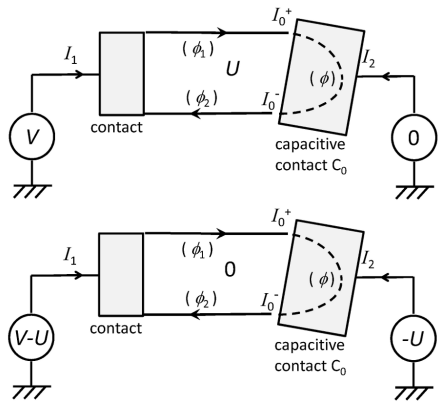,height=60mm}}}
\caption{ \textbf{One ohmic and one capacitive contact.} Top and bottom figures differ by a gauge transformation which shifts the overall potential by $U$.}
\label{FigS2}
\end{figure}

We now consider the case where contact (2) is so resistive that it can be replaced by a small but finite capacitance $C_{0}$
connected to the output of the transmission line (Figure \ref{FigS2}). The non-interacting limit is treated by setting the geometrical capacitances $C_{3}$ and $C_{4}$ to be infinite. The very short capacitive contact length (few $\mu$m) can be represented by a quantum capacitance $C_{q}=\nu \frac{e^2}{h}\tau$ in series with $C_{0}$, small enough to be in the regime where $\omega \tau \ll 1$ ($\tau\simeq$ few ps). Following \cite{SM1,SM2,SM3} the problem of the current $I=-I_{2}$ response to an ac voltage $V$ can be replaced by an equivalent problem where the internal potential $U$ in the sample is, according to gauge invariance,  subtracted off all potentials, as in the lower part of Figure \ref{FigS2}. Let $I_{0}^{+}=\nu \frac{e^2}{h}(V-U) e^{i\phi_{1}}$ be the current at the input of the capacitive contact, and $I_{0}^{-}= I_{0}^{+}e^{i\phi}$ the current at the output of the contact. $\phi_{1}=\omega \tau_{1}$ is the temporal phase accumulated by the current emitted by contact (1) up to the entrance of the capacitive contact and $\phi=\omega \tau$ the temporal phase accumulated within the capacitive contact. We have
\begin{equation}
I_{2}=-I=I_{0}^{+}(1-e^{i\phi})=(-iC_{0}\omega)\,U.
\end{equation}
For $\omega \tau \ll 1$, we find the impedance
\begin{equation}
\frac{V}{I}=\frac{1}{-iC_{0}\omega}+e^{-i\phi_{1}}\left( R_{q}+\frac{1}{-iC_{q}\omega} \right).
\end{equation}
For $\phi_{1}=0$ we recover the expression of the mesoscopic capacitor impedance of \cite{SM3,SM4}, where $R_{q}=\frac{h}{2\nu e^{2}}$. For small $\omega$ and large $C_{0}$ we have $I(t)\simeq C_{q}dV(t-\tau_{1})/dt$. We expect the dominant part of the pulse response to be the derivative of the step function pulse emitted by the generator used in the TDR mode.

\subsubsection{II. Inclusion of short range interactions}

\vspace{4mm}

\begin{figure}[h]
\vspace{2mm} \centerline{\hbox{
\epsfig{figure=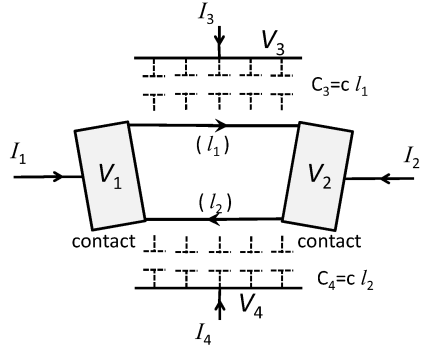,height=50mm}}}
\caption{\textbf{Two ohmic contacts.} Conductance between two ohmic contacts connected with two chiral states of lengths $l_1$ and $l_2$.}
\label{FigS3}
\end{figure}

We now introduce Coulomb interactions in the short range limit. We start with the simple case of two perfect ohmic contacts just as above in Section \textit{I}. The problem being chiral, in the absence of backscattering we expect the upper and lower edge dynamics to be fully decoupled. We follow the approach used in \cite{SM2} for modeling the ac response of a (non-chiral) 1D interacting wire. This approach treats the wire, or Tomonaga-Luttinger liquid, not in the framework of the bosonization technique but in the self-consistent random phase approximation (RPA). The Coulomb interaction is taken to be short-range. This is realized by assuming a capacitor placed very close to the electrons, see Figure \ref{FigS3}. The interaction becomes local with ac potential $u(x)=(e/c)\rho (x)$, where $\rho (x)$ is the local charge density and $c$ is the capacitance per unit length defined by the wire and the screening gate. The difference with respect to the non-chiral case enters in the polarization kernel $\Pi (x,x^{\prime})$. For the upper edge, the last right hand term of Eq. (4) in \cite{SM2}, $\frac{iq_{F}\nu_{F}}{2}e^{iq_{F}|x-x^{\prime}|}$, is here replaced by
$iq_{F}\nu_{F} e^{iq_{F}(x-x^{\prime})}$ for $x^{\prime} < x$ and by $0$ for $x^{\prime} > x$ as a result of chirality. Here we use the notation $q_{F}=\omega/v_{D}$  where the drift velocity replaces the Fermi velocity used for the non-chiral case. $\nu_{F}=c_{q}/e^{2}$ is the one-dimensional density of states with $c_{q}=\nu e^{2}/hV_{D}$ the quantum capacitance per unit length of the upper edge. Similar definitions hold for the lower edge. The solution of the self consistent equations for the propagation of the potential or the charge density along the upper edge is sought in the form of $e^{iqx}$ (while both components $e^{\pm iqx}$ are necessarily present in the non-chiral case).
We find  $q=q_{F}\frac{c}{c+c_{q}}$. As a result, the phase velocity of the, now collective, excitation describing the charge propagation becomes

\begin{equation}
v_{\phi}=\frac{\omega}{q}=\frac{\nu e^{2}}{h}\left(\frac{1}{c_{q}}+\frac{1}{c}\right)=v_{D}+v_{EMP}
\end{equation}

We recover the expression of Eq.(1) in the main text where the edge magneto-plasmon velocity in the screened regime
is now $v_{EMP}=\sigma_{xy} /c$.

Coming back to the problem of the current at contact (2) in response to the ac voltage applied at contact (1), the only modification to the expression obtained in Section \textit{I} is to replace the drift velocity by the velocity $v_{\phi}$. Specifically, in the notation of Figure \ref{FigS1}, we take $V_{2}=0$ and $V_{4}=0$, such that the two edges are decoupled. We find

\begin{equation}
I_{1}=\nu \frac{e^2}{h}
\end{equation}
\begin{equation}
I_{2}=\nu \frac{e^2}{h}\left(- V_{1}e^{iq l_{1}}  - V_{3}(1 - e^{iq l_{1}})\right)
\end{equation}
\begin{equation}
I_{3}=\nu \frac{e^2}{h}(V_{3}-V_{1})(1 - e^{iq l_{1}})
\end{equation}

We see that this is the same solution as the one given by equations (1-3) with $\phi_{1}$ replaced by $q l_{1}$. Adding the contribution of a finite $V_{2}$ and $V_{4}$, a set of equations identical to Eq.(1-4) is recovered with $\widetilde{V}_{3(4)}$ replaced by $V_{3(4)}$ (indeed, the geometrical capacitances are already taken into account by the short range interaction which renormalizes the propagation velocity). The main consequence for the pulse response is that the role of the ohmic contact is the same as in the non-interacting case. Despite the strong change in the dynamics of the current propagation, full absorption of the pulse is expected if at least one ohmic contact is present. This prevents the pulses from making several turns along the sample edge and no echo is expected, as is presently the case in our experiment.

\end{document}